\documentclass[12pt]{iopart}

\usepackage{graphicx}
\usepackage{bm}
\usepackage{bbm}
\usepackage{cite} 

\usepackage[mathlines]{lineno}
\usepackage[colorinlistoftodos]{todonotes}
\usepackage[colorlinks=true, allcolors=blue]{hyperref}

\begin{document}

\title[Dirac materials  under linear polarized light]{Dirac materials  under linear polarized light: quantum wave function evolution and topological Berry phases as classical charged particles trajectories under electromagnetic fields 
}

\author{V. G. Ibarra-Sierra$^{1, 2}$, J. C. Sandoval-Santana$^{1,3,*}$, A. Kunold$^{1}$, Sa\'ul A. 
Herrera$^{4}$, Gerardo G. Naumis$^{4}$}

\address{
$^{1}$\textit{\'Area de F\'isica Te\'orica y Materia Condensada,
Universidad Aut\'onoma Metropolitana Azcapotzalco, Av. San Pablo 180,
Col. Reynosa-Tamaulipas, 02200 Cuidad de M\'exico, M\'exico.}\\

$^{2}$\textit{Facultad de Ciencias, Universidad Aut\'onoma de Baja California, Apartado Postal 1880, 22800 Ensenada, Baja California, M\'exico.}\\

$^{3}$\textit{Centro de Nanociencias y Nanotecnolog\'ia, Universidad Nacional Aut\'onoma de M\'exico, Apartado Postal 2681, 22800 Ensenada, Baja California, M\'exico.}

$^{4}$\textit{Departamento de Sistemas Complejos, Instituto de Fisica,
Universidad Nacional Aut\'onoma de M\'exico, Apartado Postal 20-364,01000,
Ciudad de M\'exico, M\'exico.}

}

\ead{*jcss@azc.uam.mx}


\begin{abstract}

The response of electrons under linearly polarized light
in Dirac materials as borophene or graphene
is analyzed in a continuous wave regime for an arbitrary intense
field. Using a rotation and a time-dependent phase transformation,
the wave function evolution is shown to be governed by
a spinor-component decoupled Whittaker-Hill equation.
The numerical solution of these equations enables
to find the quasienergy spectrum. For borophene it reveals a strong anisotropic 
response.
By applying an extra unitary transformation, the wave functions are proven
to follow an Ince equation.
The evolution of the real and imaginary parts of the wave function is
interpreted as the trajectory of a classical charged particle
under oscillating electric and magnetic field.
The topological properties of this forced quantum system are studied
using this analogy.  In particular, in the adiabatic driving regime, the system is described with an effective Matthieu equation while in the non-adiabatic regime the full Whittaker-Hill equation is needed. From there, it is possible to separate the dynamical and Berry phase contributions to obtain the topological phase diagram due to the driving. Therefore, a different path to perturbation theory is developed to obtain time-driven
topological phases. 

\begin{description}
\item[Keyword: ]
  Dirac materials, topological phases, time driven systems, graphene, borophene
\end{description}

\end{abstract}

\section{Introduction}
The possibility of inducing a nontrivial topology in condensed matter systems by means of an electromagnetic drive has been the subject of many research works in recent years \cite{Bao2021,Lindner2011,Rudner2020,Seetharam2015,Platero2013,Lago2015,Piskunow2015,Rodriguez2021,Li2020,Vogl2020}, some of which have focused in Dirac systems such as graphene \cite{Foa2020,Sentef2015,DAlessio2015,Oka2009,Calvo2011,Calvo2013,Gonzalo2014,Piskunow2014}.

More recently, two dimensional (2D) boron allotropes, also called borophenes,
exhibiting Dirac cones have attracted attention due to their remarkable anisotropic properties \cite{Nakhaee2017,Feng2017,Zabolotskiy2016,Lherbier2016}. In a series of previous works we studied the effects of linear\cite{Champo2019} and elliptical\cite{ibarra2019dynamical,Kunold2020} polarized  electromagnetic fields in borophene.
The focus of such works was in finding
the quasienergy spectrum, the associated photo currents and induced transitions, useful to design 2D electronic devices \cite{Terrones2009}. 
In the present work, we exploit these results and others \cite{Napitu2020,Fan2018,Li2020} to study the wave function evolution and the topology induced  by the interplay between an electromagnetic drive and the electrons in a Dirac system.
To do so, we introduce a drive to the continuum model for a general tilted Dirac Hamiltonian. Then we use the Floquet formalism to establish
the equivalence between the quantum wave function evolution
and the motion of a classical charged particle in a time-dependent electromagnetic field. This is achieved through a unitary transformation
that turns the time-dependent Schr\"odinger equation into a Ince differential equation. The advantage of such formalism is that we can
use previous studies on the trajectories of particles under  time dependent electromagnetic drive \cite{lewis1968} as well as from studies of differential equations with
time-dependent coefficients \cite{Kovacic,Daniel}. In particular, the quantum spectrum is given by the stability charts of Mathieu's and Hill's equations and its generalizations \cite{Kovacic}. Such
equations occurs in electromagnetism, mechanics, cooling of ions, aerodynamics, marine research, biomedical engineering, celestial mechanics and general relativity \cite{Valluri}. 
Thus, the induced topological phases in the quantum problem are discussed at different regimes
from the perspective  of classical orbital precession.

\section{Dirac materials subject to electromagnetic fields}\label{sec:DiracHamiltonian}
In this section we address the model
and summarize previous results concerning
the Floquet theory and the quasienergy spectrum
for time-driven Dirac materials \cite{lopez2008analytic,ibarra2019dynamical,Sandoval2020}. 

\subsection{Isotropic, anisotropic and tilted Dirac materials}

The most general, low-energy Dirac Hamiltonian close to one of the Dirac
points,  is given by 
\cite{Zabolotskiy2016,verma2017effect,Champo2019,HerreraNaumisKubo2019}
\begin{equation}\label{ec:AnsotropicDiracHamiltonian}
\hat{H}= \hbar v_t k_y\hat{\sigma}_0
+\hbar\left[v_x k_x\hat{\sigma}_x +v_y k_y\hat{\sigma}_y\right],
\end{equation}
where $ k_x$ and $k_y$ are the components of the two-dimensional
momentum vector $\boldsymbol{k}$,  $\hat{\sigma}_x$
and $\hat{\sigma}_y$ are the Pauli matrices,
and $\hat{\sigma}_0$ is the $2\times 2$ identity matrix.
This Hamiltonian describes, for example, $8-Pmmn$ borophene. 
The three velocities in the anisotropic $8-Pmmn$ borophene Dirac
Hamiltonian (\ref{ec:AnsotropicDiracHamiltonian})
are given by $v_x=0.86v_F$, $v_y=0.69v_F$ and $v_t=0.32v_F$ where
$v_F=10^6\,m/s$ \cite{Zabolotskiy2016}
is the Fermi velocity.
In Eq. (\ref{ec:AnsotropicDiracHamiltonian}), the
last two terms give rise to the familiar form of the
kinetic energy leading to the Dirac cone and
the first one tilts the
Dirac cone in the $y$ direction.
These two features are contained in the
energy dispersion relation\cite{ibarra2019dynamical}
\begin{equation} \label{ec:EnergyDispersion}
E_{\eta,k}=\left(\frac{v_t}{v_y}\right) \tilde{k}_y +\nu\epsilon,
\end{equation}
where
\begin{equation}\label{ec:EpsilonCoefficient}
\epsilon_{\bm{k}}=\sqrt{\tilde{k}_{x}^{2}+\tilde{k}_{y}^{2}},
\end{equation}
and $\nu= \pm 1$ is the band index.
In Eq. (\ref{ec:EnergyDispersion}), we used the set of
renormalized moments
$\tilde{k}_x=\hbar v_x k_x$, $\tilde{k}_y=\hbar v_y k_y$.
The corresponding free electron wave function is,
\begin{equation}\label{ec:WaveFunctionFree}
\psi_{\nu}(\boldsymbol{k})
  = \frac{1}{\sqrt{2}}
  \left[
  \begin{array}{c}
     1 \\
     \nu\exp(i \theta_{\boldsymbol{k}} ) 
    \end{array} \right],
\end{equation}
where $\theta_{\boldsymbol{k}}= \tan^{-1} (\tilde{k}_y/\tilde{k}_x)$.
The case of graphene can be recovered by setting $v_t=0$ and $v_x=v_y=v_F$ and
for non-uniform strained graphene requires $v_t=0$ and $v_x \neq v_y$.

\subsection{Linearly polarized waves and Whittaker-Hill equation} 

Now we consider a charge carrier, described by the two-dimensional
anisotropic Dirac Hamiltonian, subject to an electromagnetic wave that 
propagates along a direction perpendicular to the surface of the crystal.
The effects of the electromagnetic field
are introduced in the Dirac Hamiltonian
(\ref{ec:AnsotropicDiracHamiltonian}) through the Peierls 
substitution\cite{lopez2008analytic,lopez2010graphene} 
$\hbar\boldsymbol{k}\rightarrow\hbar\boldsymbol{k}-e\boldsymbol{A}$
where $\boldsymbol{A}=(A_x,A_y)$ is the vector potential
of the electromagnetic wave.
Adopting a gauge in which $\boldsymbol{A}$ only
depends on time brings
a significant simplification.
The Hamiltonian (\ref{ec:AnsotropicDiracHamiltonian})
is thus transformed into \cite{lopez2008analytic,lopez2010graphene} ,
\begin{equation} \label{ec:DiracHamiltonianUnderElectromagneticField} 
\hat{H}= \frac{v_t}{v_y}\left(\tilde{k}_y-e v_y A_y\right)\hat{\sigma}_0
 +\left(\tilde{k}_x-e v_x A_x\right)\hat{\sigma}_x 
 +\left(\tilde{k}_y-e v_y A_y\right)\hat{\sigma}_y.
\end{equation}
Assuming the electromagnetic wave to be
linearly polarized and propagating along
the $z$ direction, the vector potential
can be written as
\begin{equation}\label{ec:VectorPotential}
\boldsymbol{A}=\frac{E_0}{\Omega}\cos(\Omega t)\boldsymbol{\hat{r}},
\end{equation}
where $\boldsymbol{\hat{r}}=(1,0)$ is the polarization vector,
$E_0$ is the uniform amplitude of the electric field
and $\Omega$ is the angular frequency of the electromagnetic wave.
It is noteworthy that the field $\boldsymbol{A}$ is not quantized and
is treated clasically. Thus, our results are only valid for
quantum coherent field composed of
a large number of photons. 
In the Schr\"odinger equation corresponding to
(\ref{ec:DiracHamiltonianUnderElectromagneticField}),
\begin{equation}\label{ec:DiracEquationOne}
i \hbar \frac{d}{d t} {\boldsymbol{\Psi}}(t)
  =\hat{H}(t){\boldsymbol{\Psi}}(t),
\end{equation}
the two dimensional spinor can be expressed as
$\boldsymbol{\Psi}(t)
 =\left(\Psi_{A}(t),\Psi_{B}(t)\right)^{\top}$, where
$A$ and $B$ label the two sublattices. Formally, the solutions can be obtained from the  time
evolution operator  $\hat{\mathcal{U}}(t)$ as,
\begin{equation}
{\boldsymbol\Psi}(t)= \hat{\mathcal{U}}(t) {\boldsymbol\Psi}(0).
\end{equation}

Due to the time periodicity of the Hamiltonian
$\hat{H}(t)=\hat{H}(t+T)$ where $T=2\pi/\Omega$, solutions must comply
with the Floquet theorem \cite{lopez2008analytic,kibis2010metal}
that states that the evolution operator must have the form
\begin{equation}
    \hat{\mathcal{U}}(t)=\exp\left(-\frac{i}{\hbar}\hat{H}_e t\right)\hat{\mathcal{W}}(t),
    \label{eq:evolfloquet}
\end{equation}
where $\hat{\mathcal{W}}(t+T)=\hat{\mathcal{W}}(t)$ and $\hat{H}_e$ is
called the effective Hamiltonian.
The eigenvalues of $\hat{H}_e$ are the quasienergies of $\hat{H}(t)$
\begin{equation}
    \mathcal{E}_{\eta,j,m}(\boldsymbol{k})
      =-\frac{\hbar\omega}{2\pi} \arg[u_{\eta,j}(\boldsymbol{k})]+m\hbar \omega,
      \label{eq:quasieneries}
\end{equation}
where $u_{\eta,j}(\boldsymbol{k})$ are the two eigenvalues of 
$\hat{\mathcal{U}}(T)$,
and $m=0,\pm 1, \pm 2,\dots$ and
$j=1,2$ denote the Floquet zone and the band 
respectively \cite{Sandoval2020}.

The main challenge in
deducing the wave function's explicit form
resides in unraveling the coupling between
the spinor components 
$\Psi_{A}(t)$ and $\Psi_{B}(t)$
that arise from terms proportional
to $\hat\sigma_x$ and $\hat\sigma_y$
in Eq. (\ref{ec:DiracHamiltonianUnderElectromagneticField}).
To uncouple the spinor components we proceed
as follows.
First, to transform the non-diagonal $\hat\sigma_x$
matrix into $\hat\sigma_z$,
we apply a $45^\circ$ rotation around the $y$ axis
of the form \cite{Sandoval2020},
\begin{equation} \label{ec:SolutionOne}
\mathbf{\Psi}(t)
=\exp\left[-i\left(\frac{\pi}{4} 
\right)\hat{\sigma}_y\right]\mathbf{\Phi}(t).
\end{equation}
Substituting (\ref{ec:SolutionOne}) into 
Eq. (\ref{ec:DiracEquationOne}) we obtain
\begin{equation}
\label{ec:Schrokxky}
i \frac{d}{d\phi}\mathbf{\Phi}(\phi)
=\frac{2}{\hbar \Omega}\left[\left(\frac{v_t}{v_y}\right) 
\tilde{k}_{y} \hat{\sigma}_0+ \tilde{\Pi}_x\hat{\sigma}_z
+ \tilde{k}_{y} \hat{\sigma}_y \right]\mathbf{\mathbf{\Phi}}(\phi) \, ,
\end{equation}
where the only off-diagonal terms
originate from $\hat\sigma_y$.
In the foregoing equation, the scaled time is defined as
$\phi= \Omega t/2$,
the scaled momentum $\tilde{\Pi}_x=\tilde{k}_x-\zeta_x \cos(2 \phi)$ and the frequency-weighted  induced  dipole  moment is,
\begin{equation}
\zeta_x=\frac{ev_xE_x}{\Omega}.
\end{equation}
The spinor components of
$\mathbf{\Phi}(\phi)=\left(\Phi_{+}(\phi),\Phi_{-}(\phi)\right)^{\top}$
are given by
$\Phi_{+}(\phi)=[\Psi_A(\phi)+\Psi_B(\phi)]/\sqrt{2}$ and 
$\Phi_{-}(\phi)=[\Psi_A(\phi)-\Psi_B(\phi)]/\sqrt{2}$.
Second, the term proportional to $\hat{\sigma}_0$
in  Eq. (\ref{ec:Schrokxky})
is removed by adding a time-dependent
phase to the wave function
\begin{equation}\label{ec:suprimdiag}
\mathbf{\Phi}(\phi)=\exp\left[-2i\left(\frac{v_t}{v_y}\right)\frac{ \tilde{k}_y}{\hbar \Omega}\, \phi\, \hat{\sigma}_0\right]\bm{\chi}(\phi),
\end{equation}
where $\bm{\chi}(\phi)=(\chi_{+1}(\phi),\chi_{-1}(\phi))^{\top}$. 
Finally, after inserting Eq. (\ref{ec:suprimdiag}) into Eq.
(\ref{ec:Schrokxky}), differentiating both sides with respect to $\phi$ and using Eq. (\ref{ec:Schrokxky}) to leave out the first order derivative, the resulting differential equation
takes the form
of a  Whittakker-Hill equation \cite{magnus2013hill,Sandoval2020}
\begin{equation}\label{ec:Hilluncoupled}
\bm{\chi}''(\phi)+\mathbbm{F}(\phi)\bm{\chi}(\phi)=0 \, , 
\end{equation}
where the matrix $\mathbbm{F}(\phi)$ is defined as
\begin{equation}
\label{ec:HillFinal}
\mathbbm{F}(\phi)=\left[a_{\bm{k}}+q_1\cos(2\phi)
+q_2\cos(4\phi)\right]\hat{\sigma}_{0}+iq_3\sin(2\phi)\hat{\sigma}_{z}.  
\end{equation}
The Hill equation parameters are defined as,
\begin{equation}
\label{ec:Hill_a}  
a_{\bm{k}}=\left(\frac{2\epsilon_{\bm{k}}}{\hbar\Omega}\right)^{2}
+2q_0^{2}, 
\end{equation}
\begin{equation}
\label{ec:Hill_q1}   
q_1=-8q_0\left(\frac{\tilde{k}_x}
{\hbar\Omega}\right), 
\end{equation}
\begin{equation}
\label{ec:Hill_q2}   
q_2=2q_0^{2},
\end{equation}
\begin{equation}
\label{ec:Hill_q3}
q_3= 4q_0,
\end{equation}
where
\begin{equation}
q_0=\frac{\zeta_{x}}{\hbar\Omega}=\frac{ev_xE_x}{\hbar \Omega^{2}},
\end{equation}
is the ratio between two characteristic energies of the system:
the electric-field-induced  dipole moment $ev_x/\Omega$ with  energy $ev_xE_x/\Omega$
and the photon energy $\hbar\Omega$\cite{Mojarro2020,ibarra2019dynamical}.
Thereby, $\epsilon/\hbar\Omega$
is the ratio of the electron kinetic energy to the photon energy,
$\zeta_{x}/\hbar\Omega$  is the ratio of
the work done on the charged carriers by the electromagnetic wave
to the photon energy and $\tilde{k}_x/\hbar\Omega$
is the ratio of the $x$ contribution of the electron kinetic energy
to the photon energy.

Expressing (\ref{ec:Hilluncoupled}) as
a second order differential equation
is quite advantageous for the calculations that follow.
First, the evolution operator that propagates 
the state $\chi(\phi)$ in time must be diagonal
since $\mathbbm{F}(\phi)$ is solely composed of the diagonal
matrices $\hat\sigma_0$ and $\hat\sigma_z$.
As a consequence of this,
the scalar differential
equations for the $\chi_{+1}(\phi)$ and $\chi_{-1}(\phi)$
spinor components decouple. Moreover, the differential
equation for the $\chi_{-1}(\phi)$ component turns out to be
the complex conjugate of the one for $\chi_{+1}(\phi)$.
Both differential equations may be summarized by
\begin{equation}
\label{ec:HillCompoentEta}
\chi_{\eta}''(\phi)+\left[a_{\bm{k}}+ q_1\cos(2\phi)+q_2\cos(4\phi)+i\eta q_3 \sin(2\phi)\right]\chi_{\eta}(\phi)=0,    
\end{equation}
where $\eta=\pm 1$.
In principle, we can obtain the solution for $\eta=-1$ 
from the  $\eta=1$ solution. This is
done by making the replacement $\phi \rightarrow -\phi$ in Eq. (\ref{ec:HillCompoentEta}), as $\eta \sin{2\phi}= \sin{2\eta \phi}$. The cosines and second derivative terms are not affected by a change of sign of $\phi$. Therefore, the solutions are related by,
\begin{equation}
\label{eq:chitimereversal}
\chi_{-1}(\phi)= \chi_{1}(-\phi).
\end{equation}
However, in general the initial conditions on the
first derivative of $\chi_{\eta}(\phi)$ are restricted by
Eq. (\ref{ec:Schrokxky}) and thus
Eq. (\ref{eq:chitimereversal}) can only be used for $\tilde{k}_y=0$.

We can also obtain a useful alternative expression to
Eq. (\ref{ec:HillCompoentEta})  by writing $\epsilon_{\bm{k}}$
in terms of the scaled moments, 
\begin{equation}
\label{ec:HillFinalalterative}
\chi_{\eta}''(\phi)+4\bigg[\left(\frac{\tilde{k}_x}{\hbar\Omega}-q_0\cos(2\phi)\right)^{2}+\left(\frac{\tilde{k}_y}{\hbar\Omega}\right)^{2}+i\eta q_0\sin(2\phi)\bigg] \chi_{\eta}(\phi)=0.
\end{equation}
In the previous equation,
or Eq. (\ref{ec:HillCompoentEta}), 
the spinor components are decoupled considerably
simplifying the computation and the quasienergy spectrum analysis.
Another gain of using this particular base is that
$\chi_+(\phi)$ and $\chi_-(\phi)$ are the probability amplitudes
of the valence and conduction bands, respectively.

\begin{figure}[t!]
\begin{center}
\includegraphics[width=0.6\textwidth]{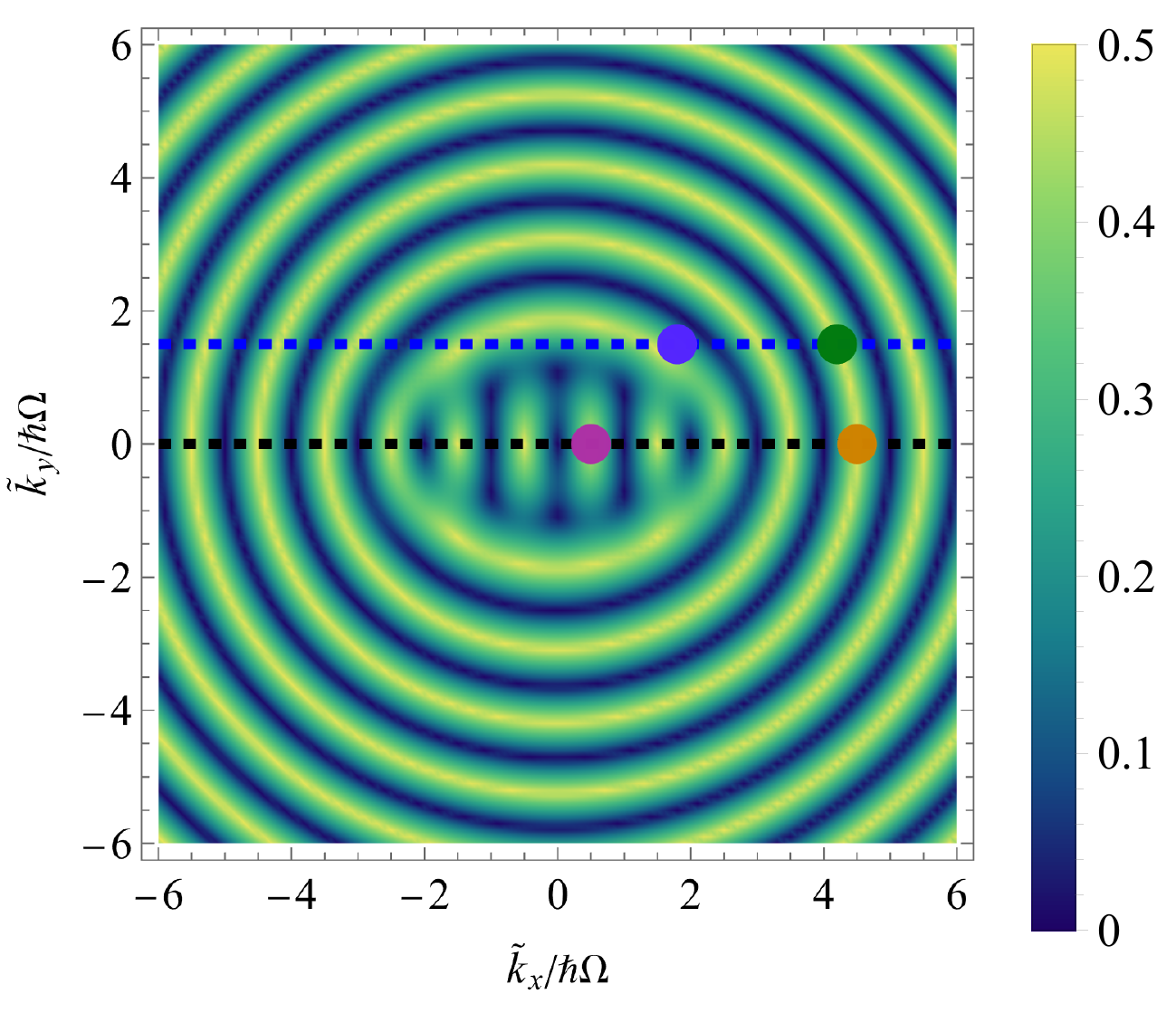}
\end{center}
\caption{\label{fig:Balon}Density plot of quasienergy spectrum $\mathcal{E}_{\eta,j,m}-v_t\tilde{k}_y/(v_y\hbar\Omega)$ for $\eta=1$ and $m=0$ as a function of $\tilde{k}_x/\hbar\Omega$ and $\tilde{k}_y/\hbar\Omega$. The amplitude and frequency of the electromagnetic wave are $E_x=4.5\,$V/m and $\Omega=50\times10^{9}\,$Hz. The horizontal dashed black and blue lines correspond to fixed values $\tilde{k}_y/\hbar\Omega=0$ and $\tilde{k}_y/\hbar\Omega=1.5$, respectively.
The dots are the states that were chosen to produce the trajectories in Figs. \ref{Fig:WittakerInceTrayec_One} and \ref{Fig:WittakerInceTrayec_Two}.}
\end{figure}

\begin{figure}[t!]
\begin{center}
\includegraphics[width=1.0\textwidth]{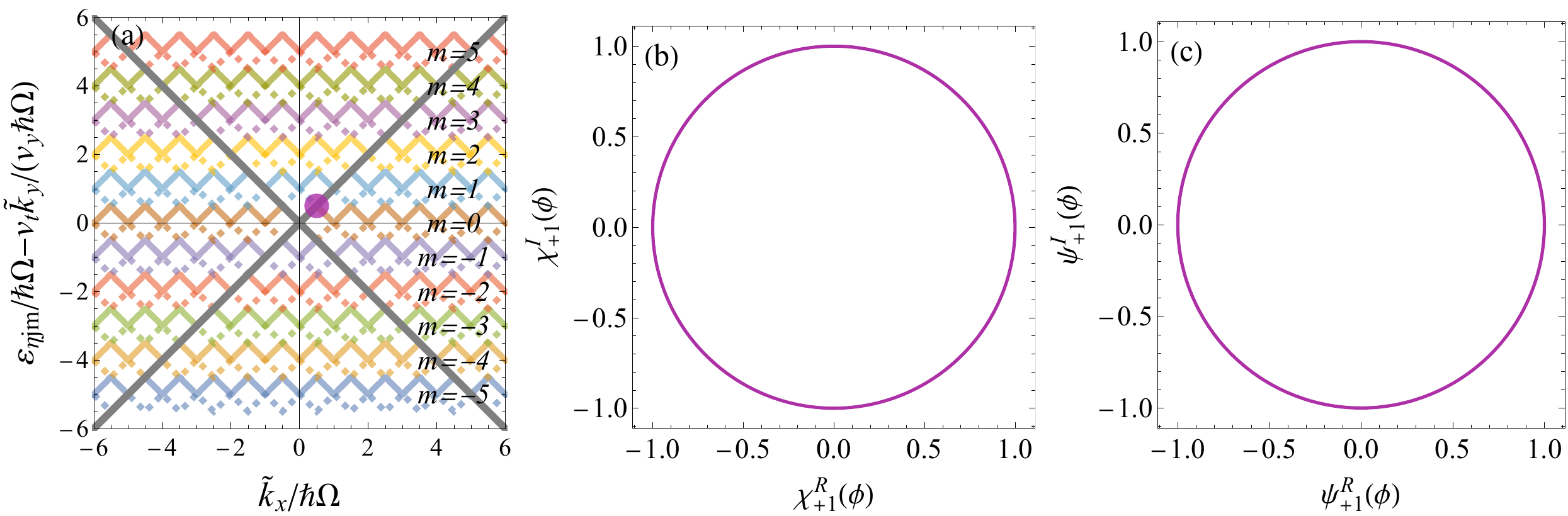}
\includegraphics[width=1.0\textwidth]{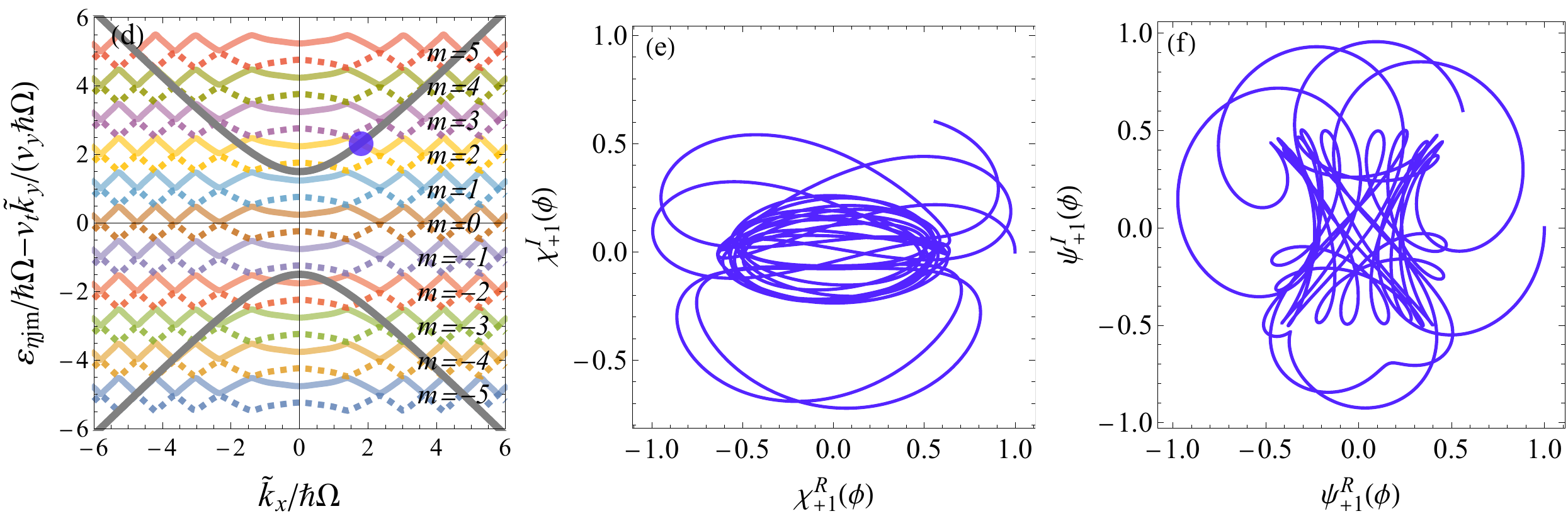}
\caption{\label{Fig:WittakerInceTrayec_One}Quasienergy spectrum $\mathcal{E}_{\eta,j,m}-v_t\tilde{k}_y/v_y\hbar\Omega$ and the Floquet quantum wave real and imaginary part for the Wittaker-Hill and Ince equations. In panels (a) and (d), we show the quasienergy spectra as a function of the momentum $\tilde{k}_x/\hbar\Omega$ using the fixed values $\tilde{k}_y/\hbar\Omega=0$ and $\tilde{k}_y/\hbar\Omega=1.5$ for $\eta=1,\,2$ and $m=0,\pm1,...,\pm5$ (Floquet zones), respectively.  In these panels, the purple dot correspond to the state where $(\tilde{k}_x/\hbar\Omega,\,\,\mathcal{E}_{\eta,j,m}-v_t\tilde{k}_y/v_y\hbar\Omega)=(0.50,\,0.50)$ and the corresponding blue dot is the state $(\tilde{k}_x/\hbar\Omega,\,\mathcal{E}_{\eta,j,m}-v_t\tilde{k}_y/v_y\hbar\Omega)=(1.79,\,2.31)$. The solid gray lines correspond to the dispersion relation given by equation (\ref{ec:EpsilonCoefficient}) for fixed $\tilde{k}_x/\hbar\Omega$ values mentioned above. In panels (b) and (c), we show the trajectories for the  Whittaker-Hill and Ince equations for $\eta=1$ and  $(\tilde{k}_x/\hbar\Omega,\,\,\mathcal{E}_{\eta,j,m}-v_t\tilde{k}_y/v_y\hbar\Omega)=(0.5,\,0.50)$, respectively. The panels (e) and (f) are the corresponding trajectories for $\eta=1$ and 
$(\tilde{k}_x/\hbar\Omega,\,\mathcal{E}_{\eta,j,m}-v_t\tilde{k}_y/v_y\hbar\Omega)=(1.79,\,2.31)$. In all these panels the value of amplitude and frequency of the electromagnetic wave are $E_x=4.85\,$V/m and  $\Omega=50\times10^{9}\,$Hz, corresponding to $q_0=2.35$. Observe how trajectories with $\tilde{k}_y/\hbar\Omega=0$ describe circles in agreement with Eq. (\ref{eq:solcirculos}).}
\end{center}
\end{figure}

\begin{figure}[t!]
\begin{center}
\includegraphics[width=1.0\textwidth]{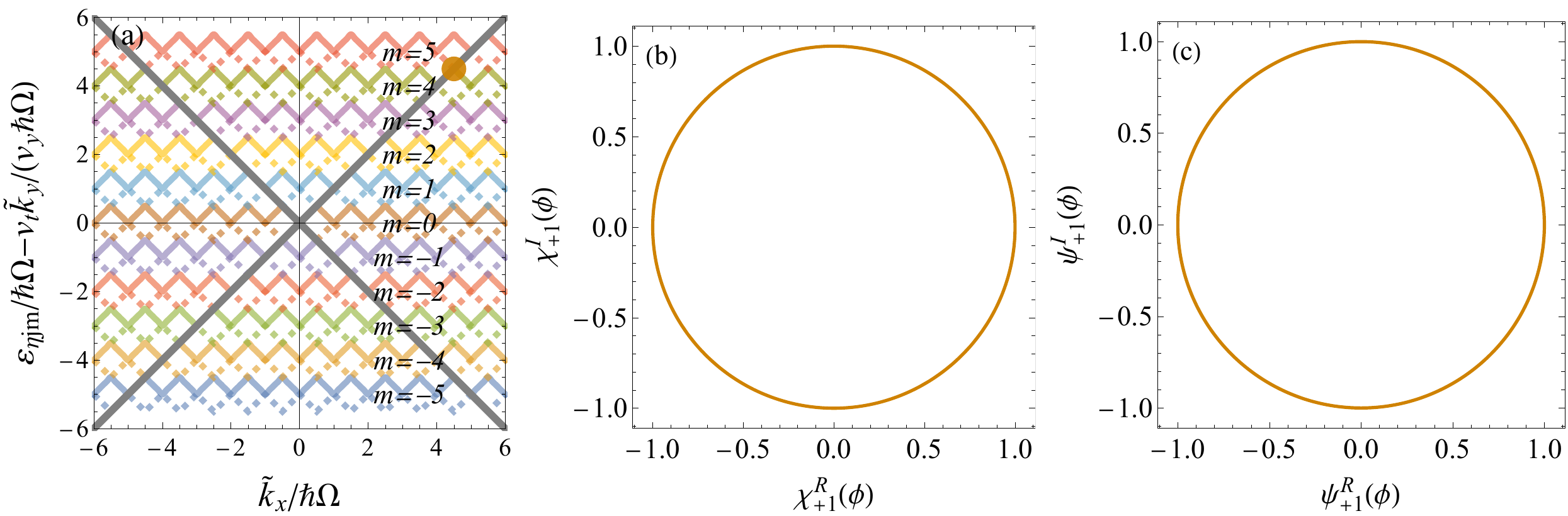}
\includegraphics[width=1.0\textwidth]{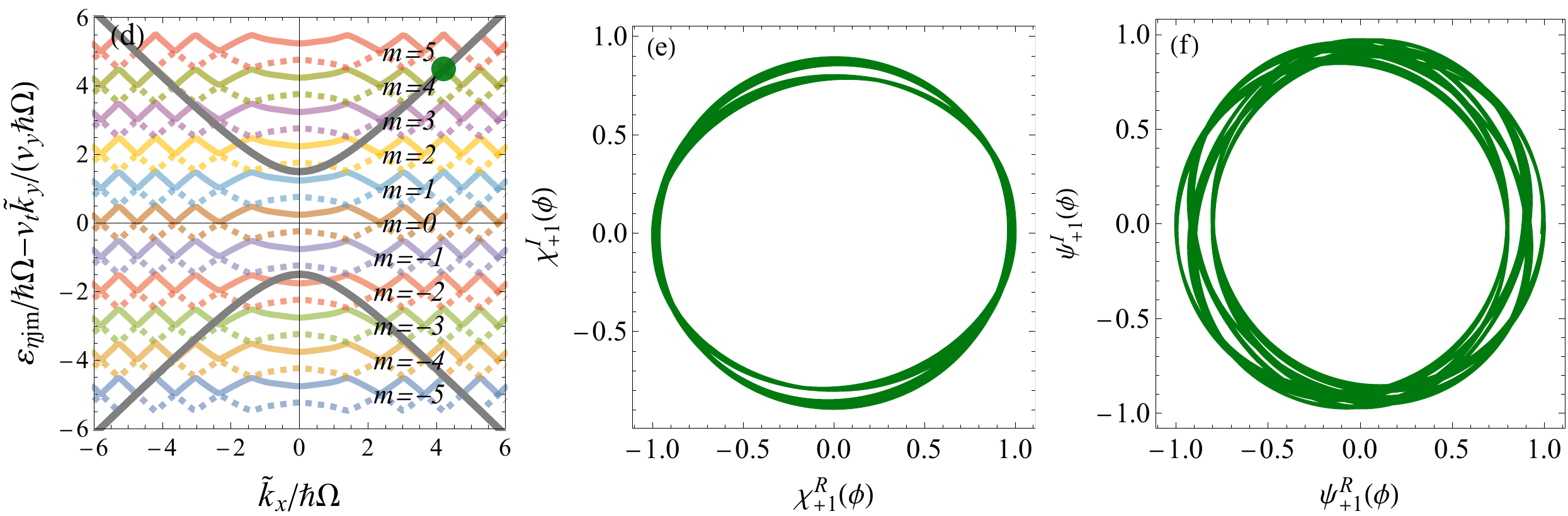}
\end{center}
\caption{\label{Fig:WittakerInceTrayec_Two}Quasienergy spectrum $\mathcal{E}_{\eta,j,m}-v_t\tilde{k}_y/v_y\hbar\Omega$ and the Floquet quantum wave real and imaginary part for the Wittaker-Hill and Ince equations. In panels (a) and (d), we show the quasienergy spectra as a function of the momentum $\tilde{k}_x/\hbar\Omega$ using the fixed values $\tilde{k}_y/\hbar\Omega=0$ and $\tilde{k}_y/\hbar\Omega=1.5$ for $\eta=1,\,2$ and $m=0,\pm1,...,\pm5$ (Floquet zones), respectively.  In these panels, the orange dot correspond to the state where $(\tilde{k}_x/\hbar\Omega,\,\,\mathcal{E}_{\eta,j,m}-v_t\tilde{k}_y/v_y\hbar\Omega)=(4.5,\,4.5)$ and the corresponding green dot is the state $(\tilde{k}_x/\hbar\Omega,\,\mathcal{E}_{\eta,j,m}-v_t\tilde{k}_y/v_y\hbar\Omega)=(4.2,\,4.49)$. The solid gray lines corresponds to the dispersion relation given by equation (\ref{ec:EpsilonCoefficient}) for fixed $\tilde{k}_x/\hbar\Omega$ values mentioned above. In panels (b) and (c), we show the trajectories for the  Whittaker-Hill and Ince equations for $\eta=1$ and  $(\tilde{k}_x/\hbar\Omega,\,\,\mathcal{E}_{\eta,j,m}-v_t\tilde{k}_y/v_y\hbar\Omega)=(4.5,\,4.5)$, respectively. The panels (e) and (f) are the corresponding trajectories for $\eta=1$ and 
$(\tilde{k}_x/\hbar\Omega,\,\mathcal{E}_{\eta,j,m}-v_t\tilde{k}_y/v_y\hbar\Omega)=(4.2,\,4.49)$. In all these panels the value of amplitude and frequency of the electromagnetic wave are $E_x=4.85\,$V/m and  $\Omega=50\times10^{9}\,$Hz, corresponding to $q_0=2.35$. Observe how trajectories with $\tilde{k}_y/\hbar\Omega=0$ describe circles in agreement with Eq. (\ref{eq:solcirculos}).}
\end{figure}

In figure \ref{fig:Balon} we present  the corresponding quasienergy spectrum
obtained using a very powerful method developed in a previous work: the monodromy 
matrix method \cite{Kunold2020} for the CB $\eta=1$ and the Floquet 
zone $m=0$. This is equivalent to finding the stability regions of Eq. 
(\ref{ec:HillCompoentEta}).  Notice  the two different regions in the spectrum.
One is at the center where an American football ball-like shape is seen.
The other, in the outer regions of the spectrum,
exhibits concentric circles.
Inside the ball, the energy of electrons is less than the energy
of a photon and corresponds to a strong electromagnetic field regime.
Outside the ball, the system can be considered in a weak interaction
regime where the quasienergy spectrum is almost similar to the non-perturbed
energy dispersion. In this case, the driving can be considered adiabatic.
The condition that defines the crossover between the adiabatic and
non-adiabatic regimes is thus  given by the perimeter of the ball, i.e.,
\begin{equation}\label{eq:transitioncond}
\epsilon_{\bm{k}} \approx \zeta_{x}=\hbar \Omega q_0.
\end{equation}
Let us then consider three 
different regimes: the adiabatic ($ev_xE_x/\hbar \Omega^2<1$),
the non-adiabatic ($ev_xE_x/\hbar \Omega^2>1$)
and the transitional ($ev_xE_x/\hbar \Omega^2\approx 1$) regimes.
In Figs. \ref{Fig:WittakerInceTrayec_One}(b),
\ref{Fig:WittakerInceTrayec_One}(e),
and \ref{Fig:WittakerInceTrayec_Two}(b),
we present the evolution of the real and imaginary parts 
of the first spinor component $\chi_{1}(\phi)$ for states chosen at 
different points of the Borophene's quasienergy spectrum.
These states, indicated with dots in Fig. \ref{fig:Balon},
correspond to the four most representative cases
listed below:
\begin{enumerate}
     \item Adiabatic regime with $\tilde{k}_y \ne 0$ (green dot),
     \item Non-adiabatic regime with $\tilde{k}_y=0$ (purple dot),
     \item Non-adiabatic regime with $\tilde{k}_y \ne 0$ (blue dot),
      \item Transitional regime with $\tilde{k}_y=0$ (orange dot).
\end{enumerate}

The trajectories seen in Figs. \ref{Fig:WittakerInceTrayec_One}(b),
\ref{Fig:WittakerInceTrayec_One}(e), \ref{Fig:WittakerInceTrayec_Two}(b), and 
\ref{Fig:WittakerInceTrayec_Two}(e) were obtained from a numerical simulation
made  by solving Eq. (\ref{ec:HillCompoentEta}).
As the initial state, we chose a band
eigenstate of the time-independent problem.
Since (\ref{ec:HillCompoentEta}) is a second order differential equation,
it requires an additional initial condition that comes
from the first derivative of the wave function in
Eq. (\ref{ec:Schrokxky}).
In Figs.  \ref{Fig:WittakerInceTrayec_One}(a),
\ref{Fig:WittakerInceTrayec_One}(d),
\ref{Fig:WittakerInceTrayec_Two}(a)
and in \ref{Fig:WittakerInceTrayec_Two}(e) we 
indicate the chosen states in several quasi spectrum
cross sections. The Floquet zone replicas are also tagged.

From Figs. \ref{Fig:WittakerInceTrayec_One}(b),
\ref{Fig:WittakerInceTrayec_One}(e), \ref{Fig:WittakerInceTrayec_Two}(b), and 
\ref{Fig:WittakerInceTrayec_Two}(e), we observe that as the wave functions
evolve their real and imaginary parts may either move along complex paths
or even describe simple circular trajectories. 
In particular, in Figs. \ref{Fig:WittakerInceTrayec_One}(b) and  
\ref{Fig:WittakerInceTrayec_Two}(c),
the real and imaginary parts of the wave function
describe circular paths. 
This can be understood from Eq. (\ref{ec:HillFinalalterative}),
or more directly from Eq. (\ref{ec:Schrokxky}),
as in this case $\tilde{k}_y=0$ the spinor components are decoupled
right from the beginning and there is no need to go into the second
derivative calculation.
Thus, $\mathbf{\chi}(\phi)=(\chi_{+1}(\phi),\chi_{-1}(\phi))$
and the solution is, 
\begin{equation}\label{eq:solcirculos}
\chi_{\eta}(\phi)= \chi_{\eta}(0)\exp\left[-2i\eta\int_0^{\phi}
  \left(q_0\cos(2\phi')-\frac{\tilde{k}_x}{\hbar\Omega}\right)d\phi'\right],
\end{equation}
or using that in this case $\epsilon_{\bm{k}}=\tilde{k_x}$, we obtain that,
\begin{equation}\label{eq:solcirculos2}
\chi_{\eta}(t)= \chi_{\eta}(0)\exp\left[-i\eta\left(q_0 \sin{(2\Omega t)}-\frac{2\epsilon_{\bm{k}} t}{\hbar}\right)\right].
\end{equation}
The real and imaginary parts of this solution describe the
circular paths  shown in Figs.  \ref{Fig:WittakerInceTrayec_One}(b) and  
\ref{Fig:WittakerInceTrayec_Two}(b).
What is remarkable here is that Eq. (\ref{eq:solcirculos2}) holds in
the non-adiabatic, the adiabatic and the transitional regime,
i.e., it covers items 2 and 4 of the list of most representative cases.
As we will see in the following section, this allows to
characterize the Berry phase in a simple way.

Now let us return to the case where
in Eq. (\ref{ec:Hill_q1}) 
$\tilde{k}_x \ne 0$ and $\tilde{k}_y \ne 0$
but  $q_1\gg q^{2}_0$.
Under these conditions the solution of Eq. (\ref{ec:Hill_q1}) 
requires setting $\tilde{k}_x \gg q_0\hbar \Omega$
from where $\epsilon \gg\hbar \Omega q_0$.
According to Eq. (\ref{eq:transitioncond}) such a case corresponds
to states well inside the adiabatic regime.
Using Eq. (\ref{ec:Hill_q2}) we neglect $q_2$
and Eq. (\ref{ec:HillFinalalterative}) takes the form
\begin{equation}\label{ec:HillCompoentEta_Two}
\chi_{\eta}''(\phi)+\left[a_{\bm{k}}+ q_1\cos(2\phi)+i \eta q_3 \sin(2\phi)\right]\chi_{\eta}(\phi)=0.  
\end{equation}
This is a generalized Matthieu equation \cite{Kovacic},
but if we further assume that  $\tilde{k}_x\gg\hbar \Omega/2$ and
therefore photons are far from inducing transitions, 
then Eq. (\ref{ec:HillCompoentEta})  
transforms into a simple Matthieu equation as $q_1\gg q_0$
giving
\begin{equation}\label{ec:MathieuCompoentEta}
\chi_{\eta}''(\phi)+\left[a_{\bm{k}}+ q_1\cos(2\phi)\right]\chi_{\eta}(\phi)=0.   
\end{equation}
This very well known equation describes  a  pendulum with time-driven variable 
length, or alternatively, an harmonic oscillator with natural frequency 
$\sqrt{a}a_{\bm{k}}$ with a periodically perturbed time-dependent
spring constant variation $-q_1\cos(2\phi)$.
The physical relevant solutions are given by a stability chart in the
$a$ and $q_1$ parameter space, divided in forbidden and allowed 
regions\cite{Kovacic}.   
Resonances appear around $\sqrt{a_{\bm{k}}}=n$ with $n$ integer. In this quantum 
context, the forbidden regions correspond to the spectral gaps as they represent 
non-physical runaway solutions. Each resonance defines the limit of the Floquet 
zone. Including the initial conditions, the solutions are given by,
\begin{eqnarray}\label{eq:SolucionGeneral}
\chi_{-}(\phi) &=&
\frac{\mathcal{C}\left(a_{\bm{k}},-\frac{q_1}{2},\phi\right)}
{\mathcal{C}\left(a_{\bm{k}},-\frac{q_1}{2},0\right)}
-\frac{2i\left(\tilde{k}_x-q_0\right)
\mathcal{S}\left(a_{\bm{k}},-\frac{q_1}{2},\phi\right)}
{\Omega\mathcal{S}^{\prime}\left(a_{\bm{k}},-\frac{q_1}{2},0\right)},\\
\chi_{+}(\phi) &=&\frac{2\tilde{k}_y
\mathcal{S}\left(a_{\bm{k}},-\frac{q_1}{2},\phi\right)}
{\Omega\mathcal{S}^{\prime}\left(a_{\bm{k}},-\frac{q_1}{2},0\right)}.
\end{eqnarray}
where $\mathcal{C}(a_{\bm{k}},-q_1/2,\phi)$ and 
$\mathcal{S}(a_{\bm{k}},-q_1/2,\phi)$ are the Mathieu cosine and sine functions, 
respectively. The first derivatives of the Mathieu functions are
$\mathcal{C}^\prime(a_{\bm{k}},-q_1/2,\phi)
=d/d\phi \mathcal{C}(a_{\bm{k}},-q_1/2,\phi)$ and
$\mathcal{S}^\prime(a_{\bm{k}},-q_1/2,\phi)
=d/d\phi \mathcal{S}(a_{\bm{k}},-q_1/2,\phi)$.
Aside from the initial conditions ensued by
Eq. (\ref{ec:Schrokxky}),
in the previous equations we have also assumed that
$\chi_+(0)=0$ and $\chi_-(0)=1$ which means that
initially the electron is in the valence band.
The quasienergies are obtained by using the Fourier expansion of such 
functions. This is a very interesting result as in the adiabatic regimen we
have both the oscillator's fundamental frequency and the oscillating
drive's frequency.
As we will discuss, our results must be  akin to those  of Thouless  concerning 
adiabatic phases and thus Berry topology. However, here we arrived to such result 
not by perturbation theory but from a non-perturbative approach. Moreover, our 
generalized Hill equation (\ref{ec:HillCompoentEta}) allows to find extra 
contributions  and explore non-adiabatic regimes.

Also, the previous analysis of particular cases opens the question whether if for 
$q_3\neq 0$ we can develop a suitable classical analogy to the quantum equations. 
This is the subject of the following section.

\section{Ince equation:  quantum wave functions as classical trajectories under electromagnetic fields}

The previous section showed how a quantum wave function evolution of a time-driven system can be described in some particular cases by a simple classical problem. In this section we show how to fully extend the analogy to a classical system. In particular, we study the trajectories that arise when the Whittaker-Hill equation (\ref{ec:Hilluncoupled}) is transformed into the Ince equation. Consider the following unitary transformation
\begin{equation}\label{ec:UnitTras_WhitakkerToInce}
\chi_{\eta}(\phi)=\exp\left[i\eta q_0 \sin(2\phi)\right]\psi_\eta(\phi),
\end{equation}
Substituting this last expression into Eq. (\ref{ec:Hilluncoupled}), turns the Wittaker-Hill equation into a Ince equation for $\psi_\eta(\phi)$, i.e.,
\begin{equation}\label{ec:InceEquation}
\psi''_\eta(\phi)+i f_\eta(\phi) \psi'_\eta(\phi)+g(\phi)\psi_\eta(\phi)=0,
\end{equation}
where 
\begin{equation}
f_\eta(\phi)=\eta|q_3|\cos\left(2\phi\right),
\end{equation}
\begin{equation}
g(\phi)=\left(\frac{2\epsilon_{\bm{k}}}{\hbar\Omega}\right)^{2}+q_1\cos(2\phi). 
\end{equation}

There are two advantages of the transformation
given by Eq. (\ref{ec:UnitTras_WhitakkerToInce}). One is evident by comparing with Eq. (\ref{eq:solcirculos}), as it separates the contribution that comes from $\tilde{k}_x$ whenever $\tilde{k_y}=0$. But what is more important here is the possibility of finding 
a  suitable classical analogy. Although this can be done in the Whitaker-Hill equation, the resulting fields are far from
simple known physical cases. The solution of the Ince differential equation can be decomposed into its real part and its imaginary part, that is, $\psi_{\eta}(\phi)=\psi_{\eta}^{R}(\phi)+i\psi_{\eta}^{I}(\phi)$, and we find the following set of coupled differential equations
\begin{equation}
\frac{d^2}{d\phi^2}\psi_{\eta}^{R}(\phi)-f_\eta(\phi)\frac{d}{d\phi}\psi_{\eta}^{I}(\phi)+g(\phi)\psi^{R}_\eta(\phi)=0,
\end{equation}
\begin{equation}
\frac{d^2}{d\phi^2}\psi_{\eta}^{I}(\phi)+f_\eta(\phi)\frac{d}{d\phi}\psi_{\eta}^{R}(\phi)+g(\phi)\psi^{I}_\eta(\phi)=0.
\end{equation}

 Let us know explore the classical analogy. We propose the following replacement $\psi^{R}_\eta(\phi)\rightarrow X(\phi)$, $\psi^{I}_\eta(\phi)\rightarrow  Y(\phi)$. Notice that to keep the derivation simple, we drop $\eta$ and then quote the result for $\eta=-1$ at the end of the calculation. The resulting Ince equations are written as,
 
\begin{equation}\label{ec:CoupledDiffEqs_One}
\frac{d^2 X(\phi)}{d\phi^2}-f_\eta(\phi)\frac{d}{d\phi}Y(\phi)+g(\phi)X(\phi)=0,
\end{equation}
\begin{equation}
\label{ec:CoupledDiffEqs_Two}
 \frac{d^2 Y(\phi)}{d\phi^2}+f_\eta(\phi)\frac{d}{d\phi}X(\phi)+g(\phi)Y(\phi)=0. 
\end{equation}

Consider the problem of classical particle with mass $m$ and charge $Q$ moving in a plane under an electromagnetic field given by a radial time dependent electric field,
\begin{equation}
\mathbf{E(t)}=-\frac{g(t)}{Q}\mathbf{r},\\
\end{equation}
with $\mathbf{r}=(X,Y,0)$ the position vector and a perpendicular time dependent magnetic field, 
\begin{equation}
\label{ec:MagneticPotential}
\mathbf{B(t)}=\frac{f(t)}{Q} \mathbf{\hat{k}},
\end{equation}
with $\mathbf{\hat{k}}=(0,0,1)$. The classical particle equation of motion is,
\begin{equation}
m\frac{d^2 \bm{r}}{dt^2}=Q\bm{E}(t)+Q\bm{v}\times \bm{B}(t)\label{ec:EqsMotionParticle_X},
\end{equation}
with $\bm{v}=d\bm{r}/dt$. The equation for the other valley is obtained by reversing the direction of the magnetic field in the $-\mathbf{\hat{k}}$ direction.

Comparing the set of differential equations (\ref{ec:CoupledDiffEqs_One}) and (\ref{ec:CoupledDiffEqs_Two})  arising from Ince equation with the equation (\ref{ec:EqsMotionParticle_X}) of a charged particle motion problem we see that they are similar. Now one can compare with previous literature concerning time-variable electromagnetic fields \cite{lewis1968} and visualize the phase  of the wave function
as classical trajectories. Also, we can ask what is the relationship between such trajectories and the topological properties of the wave function phases. This is the subject of the following section.


\section{Topological phases of  Dirac systems under linearly polarized light}

As is well known, the initial spark in the study of topological phases was the discovery by Berry \cite{Berry1984} that a quantum system subjected to an adiabatic change in its parameters gets a geometrical phase $\gamma_B$ in the wave function evolution, known as the Berry phase \cite{Vanderbilt2018,Xiao2010}. This phase, which must be $\gamma_B=2\pi n$ with $n$ an integer when a closed   path is made with the parameters, adds to the dynamical phase determined by the instantaneous eigenvalues $\overline{\epsilon}_{\bm{k}}(\bm{A}(t))$ of the  Hamiltonian, i.e., the total wave-function is,
\begin{equation}
\label{eq:Berrydef}
\psi_{\bm{k}}(t)=\exp{\left[i\gamma_B(t)\right]}\exp{\left[-\frac{i}{\hbar}\int_{0}^{t}\overline{\epsilon}_{\bm{k}}(t')dt'\right]}|\bm{\bm{k}}(\bm{A}(t)) \rangle,
\end{equation}

where $|\bm{k}(\bm{A}(t)) \rangle$ is an instantaneous eigenvector which satisfies,
\begin{equation}
    \hat{H}(t)|\bm{k}(t)\rangle=\overline{\epsilon}_{\bm{k}}(\bm{A}(t)) |\bm{k}(\bm{A}(t)) \rangle,
\end{equation}
and $\gamma_B(t)$ is the geometrical phase at time $t$ such that,
\begin{equation}
    \gamma_B=\gamma_B(T)-\gamma_B(0).
\end{equation}
The argument $\bm{A}(t)$ appears here to  highlight the parameter that performs the external driving. 
Therefore, the total phase obtained by the wavefunction after a one-cycle drive is,
\begin{equation}
    \gamma_{\bm{k}}(T)=\gamma_B+\gamma_D(T),
\end{equation}
where we defined the dynamical phase as,
\begin{equation}
    \gamma_D(t)=\int_{0}^{t}\overline{\epsilon}_{\bm{k}}(t')dt'.
\end{equation}
In our system, 
\begin{equation}
\overline{\epsilon}_{\bm{k}}(\phi)=\pm \sqrt{\left(\frac{\tilde{k}_x}{\hbar\Omega}-q_0\cos(2\phi)\right)^{2}+\left(\frac{\tilde{k}_y}{\hbar\Omega}\right)^{2}}.
\end{equation}

As Eq. (\ref{ec:HillFinalalterative}) allows to find the total wave function and $\gamma_D(t)$ is easy to find, is clear that in principle we can recover the Berry phase from $\gamma_B= \gamma_{\bm{k}}(T)-\gamma_D(T)$. Although such definition works for adiabatic and non-adiabatic cases, in the non-adiabatic case the phase is not necessarily geometric. However, in the adiabatic regime $\gamma_B(t)$ coincides with the usual 
definition of Berry phase.

Consider as an example the case $k_y=0$. The analytical solution Eq. (\ref{eq:solcirculos}) can be written as,

\begin{equation}\label{eq:solcirculos3}
\chi_{\pm 1}(t)= \chi_{\pm 1}(0)\exp{\left[- \frac{i}{\hbar} \int_0^{t}\overline{\epsilon}_{\bm{k}}(t')dt'\right]}.
\end{equation}

By comparing with Eq. (\ref{eq:Berrydef}), we conclude that $\gamma_B=0$ implying that in the line $k_y=0$ the system is topologically trivial. This explains why in a previous work, it was found numerically that states at such line do not have transitions to other states, even at very high fields \cite{Kunold2020}.
Mathematically, such a result follows from a special condition that the instantaneous eigenvalues satisfy for $k_y=0$,
\begin{equation}
\frac{d\overline{\epsilon}_{\bm{k}}(\phi)}{d\phi}=2q_0 \sin{2\phi},
\end{equation}
and from where the Whittaker-Hill equation is,
\begin{equation}
\chi''(\phi)+\bigg[4\overline{\epsilon}_{\bm{k}}^{2}(\phi)-i 2\eta \frac{d\overline{\epsilon}_{\bm{k}}(\phi)}{d\phi}\bigg] \chi(\phi)=0.
\end{equation}
Meanwhile, from  Eq. (\ref{eq:Berrydef}) is clear that the classical trajectories are circular in the Whittaker-Hill and Ince pictures, as in this case the resulting equations are similar. Therefore, topologically trivial phases are given by circular trajectories. Intuitively, such result is to be expected as for trivial topological phases there is not a sort of  ``phase leaking" to higher energy states. Physically, as linear polarized light is made from a superposition of the same amount of left and right photon polarization, due to momentum conservation, transitions are forbidden if the electron does not have momentum in a perpendicular direction to $\bm{A}(\bm{r})$. 

Now consider the adiabatic regime for $k_y \ne 0$.
Such conditions corresponds to item (i) of the list of most representative cases
and to the green dot in the quasi-energy spectrum of Fig. \ref{fig:Balon}.
In this limit, the solution is determined by the pure Mathieu equation Eq. (\ref{ec:MathieuCompoentEta}) and the initial conditions. Following Thouless \cite{Thouless1983}, we consider that at $t=0$ the system is in a pure valence band state of the stationary problem. According to Eq. (\ref{eq:SolucionGeneral}), this is translated into $\chi_+(0)=0$ and $\chi_-(0)=1$. However, for $k_y \ne 0$ we see from Eq. (\ref{eq:SolucionGeneral}) that $\chi_+(t) \ne 0$ for $\phi \ne 2\pi s+1$ with $s$ as an integer. As a consequence, there is a small projection onto the conduction band. This is the hallmark of a topological phase. Here we remark that Eq. (\ref{eq:SolucionGeneral}) is consistent with the topological trivial case $k_y=0$ as $\chi_+(t)=0$ at all times. Finally, for the non-adiabatic $k_y \ne 0$ case (blue dot state), the classical trajectories are far from circles as the conduction band component is not small and in fact transitions are induced. Moreover, the full Whittaker-Hill equation is needed as in this case the term $q_2$, which corresponds to a doubling of the drive frequency, will dominate. Not surprisingly, the same equation
appears in the calculation of celestial bodies orbital precession \cite{Valluri}.

\section{Conclusions}\label{sec:Conclusions}

The time evolution of an electron in a 2D Dirac material driven by
 linear polarized electromagnetic fields was found using Floquet theory. In particular, a
 set of Whittaker-Hill equations  was obtained for the bispinor wave function evolution. The resulting trajectories for the phases were obtained numerically and in some cases, it was possible to compare with the analytical results. To further decouple the bispinor components, a unitary transformation was used. This transformation allows to decouple the bispinors into an Ince equation.
 In this manner it is possible to describe the phase of the electron wave function as a classical charged particles in a time-driven electromagnetic field. Circular trajectories were identified as trivial topological phases. This
 occurs when the electron momentum is aligned with the photon momentum and transitions are thus forbidden. When this is not the case, non-trivial topological phases were identified as trajectories with a small precession. They are described by a Mathieu equation where the precession is due to a phase leaking into the conduction band states. In the non-adiabatic regime, the trajectories are complicated except for the case in which
 electrons and photons have parallel momentum. Such
 result is due to the role played by a frequency
 component that doubles the original driving frequency.

\section{Acknowledgements}\label{sec:Acknowledgements}

This work was supported by DCB UAM-A grant numbers
2232214 and 2232215, and UNAM DGAPA PAPIIT IN102620, and CONACyT project 1564464. J.C.S.S. and V.G.I.S acknowledge the total support from 
Estancias Posdoctorales por M\'exico 2021 CONACYT.\\

\bibliographystyle{unsrt}
\bibliography{biblio.bib}

\end{document}